\documentclass[pss]{wiley2sp} 
\usepackage{amsmath}
\usepackage{bm}
\usepackage[T1]{fontenc}
\usepackage[utf8x]{inputenc}
\usepackage[colorlinks=true,pdfstartview=FitV,linkcolor=black,citecolor=black,urlcolor=blue]{hyperref}
\usepackage{graphicx}
\usepackage{amsfonts}
\usepackage{amssymb}
\usepackage{mathtools}
\usepackage{cite}

\begin{document}

\newcommand{\be}{\begin{equation}}
\newcommand{\ee}{\end{equation}}
\def\ud{{\mathrm{d}}}
\def\im{{\mathrm{i}}}
\def\ex{{\mathrm{e}}}
\newcommand{\ket}[1]{| #1 \rangle}
\newcommand{\bra}[1]{\langle #1 |}
\renewcommand\Im{\operatorname{Im}}
\renewcommand\Re{\operatorname{Re}}
\def\unity{\mbox{\boldmath $1$}}
\def\a{\alpha}
\def\b{\beta}
\def\g{\gamma}
\def\d{\delta}
\def\eps{\epsilon}
\def\vareps{\varepsilon}
\def\tb{\bar{t}}
\def\heff{h_{\text{eff}}}
\def\heffb{\bar{h}_{\text{eff}}}
\newcommand\varmp{\mathbin{\vcenter{\hbox{%
  \oalign{\hfil$\scriptstyle-$\hfil\cr
          \noalign{\kern-.3ex}
          $\scriptscriptstyle({+})$\cr}%
}}}}

\title{Adiabatic Preparation of a Correlated Symmetry-Broken Initial State with the Generalized Kadanoff--Baym Ansatz}

\author{%
  Riku Tuovinen\textsuperscript{\Ast,\textsf{\bfseries 1}},
  Denis Gole\ifmmode \check{z}\else \v{z}\fi{}\textsuperscript{\textsf{\bfseries 2}},
  Michael Sch{\"u}ler\textsuperscript{\textsf{\bfseries 2}},
  Philipp Werner\textsuperscript{\textsf{\bfseries 2}},
  Martin Eckstein\textsuperscript{\textsf{\bfseries 3}},
  Michael A. Sentef\textsuperscript{\textsf{\bfseries 1}}}

\mail{e-mail
  \textsf{riku.tuovinen@mpsd.mpg.de}}

\institute{%
  \textsuperscript{1}\,Max Planck Institute for the Structure and Dynamics of Matter, 22761 Hamburg, Germany\\
  \textsuperscript{2}\,Department of Physics, University of Fribourg, 1700 Fribourg, Switzerland\\
  \textsuperscript{3}\,Department of Physics, University of Erlangen--N{\"u}rnberg, 91058 Erlangen, Germany}


\abstract{\bf%
A fast time propagation method for nonequilibrium Green's functions based on the generalized Kadanoff--Baym Ansatz (GKBA) is applied to a lattice system with a symmetry-broken equilibrium phase, namely an excitonic insulator. The adiabatic preparation of a correlated symmetry-broken initial state from a Hartree--Fock wave function within GKBA is assessed by comparing with a solution of the imaginary-time Dyson equation. We find that it is possible to reach a symmetry-broken correlated initial state with nonzero excitonic order parameter by the adiabatic switching procedure. We discuss under which circumstances this is possible in practice within reasonably short switching times.
}

\maketitle   

\section{Introduction}\label{sec:intro}
A standard approach to nonequilibrium many-body problems is the nonequilibrium Green's function (NEGF) technique~\cite{Danielewicz1984,svlbook,Balzer2013book}, where dynamical information about the studied system, e.g. electric currents or the photoemission spectrum, is encoded into the Green's function. To access this information, we have to consider the coupled integro-differential equations of motion for the Green's function, the Kadanoff--Baym equations~\cite{Baym1961,kbbook}, whose efficient solution is far from trivial due to the double-time structure~\cite{Semkat1999,Kohler1999,Kwong2000,Dahlen2006,Dahlen2007,Stan2009,Schueler2018mem}. The Generalized Kadanoff--Baym Ansatz (GKBA) offers a simplification by reducing the two-time-propagation of the Green's function to the time-propagation of a time-local density matrix~\cite{Lipavsky1986}. This computational advantage brought by the GKBA has been realized and broadly applied in many contexts, such as quantum-well systems~\cite{Jahnke1997,Balzer2012,Hermanns2012}, molecular junctions~\cite{Galperin2008,Ness2011,Latini2014}, metallic clusters~\cite{Pal2009}, Hubbard nanoclusters~\cite{Balzer2013,Hermanns2014,Schluenzen2017}, and pump-probe spectroscopies for atomic and molecular systems~\cite{Perfetto2015a,Perfetto2015b,Perfetto2018}.

For a full two-time Green's function calculation, a correlated initial equilibrium state for the dynamics can be found by the extended imaginary-time-contour~\cite{Ku2002,Dahlen2005}. For the GKBA, however, no corresponding equilibrium approximation is known~\cite{Hermanns2012}. Instead, it has been customary to use the adiabatic theorem to ``switch-on'' the many-body effects adiabatically. In this adiabatic switching (AS) procedure the time-propagation {\`a} la GKBA is started from a noninteracting or a mean-field initial state, such as a Hartree--Fock initial state obtained from a separate calculation. The many-body self-energies are then slowly switched on according to a suitably chosen ramp function, and the system is evolved to a correlated equilibrium state. This method has so far proven successful in preparing the correlated equilibrium state, but to the best of our knowledge the AS procedure has not been attempted for systems with a symmetry-broken initial state, such as superconducting~\cite{BCS1,BCS2,Gorkov1959,Kemper2015,Sentef2016,Babadi2017,Murakami2017sc} or excitonic insulator~\cite{Mott1961,Keldysh1965,Jerome1967,Golez2016,Matveev2016,Murakami2017,Freericks2017,Tanaka2018} phases. It is the purpose of this paper to assess the validity and accuracy of the GKBA with the AS procedure for a prototypical symmetry-broken system of an excitonic insulator. To this end, we study a simple model of a one-dimensional two-band system with interband Hubbard interaction~\cite{Golez2016}.

The paper is organized as follows. We introduce the model system in Sec.~\ref{sec:ei}. In Sec.~\ref{sec:background} we outline the main points of the underlying NEGF theory, together with some details on the implementation of the GKBA. The adiabatic preparation of symmetry-broken initial states by the GKBA is shown and analyzed in Sec.~\ref{sec:results}. Finally, in Sec.~\ref{sec:concl} we draw our conclusions and discuss future prospects.

\section{Excitonic insulator}\label{sec:ei}
\begin{figure}[t]
\centering
\includegraphics[height=0.25\textwidth]{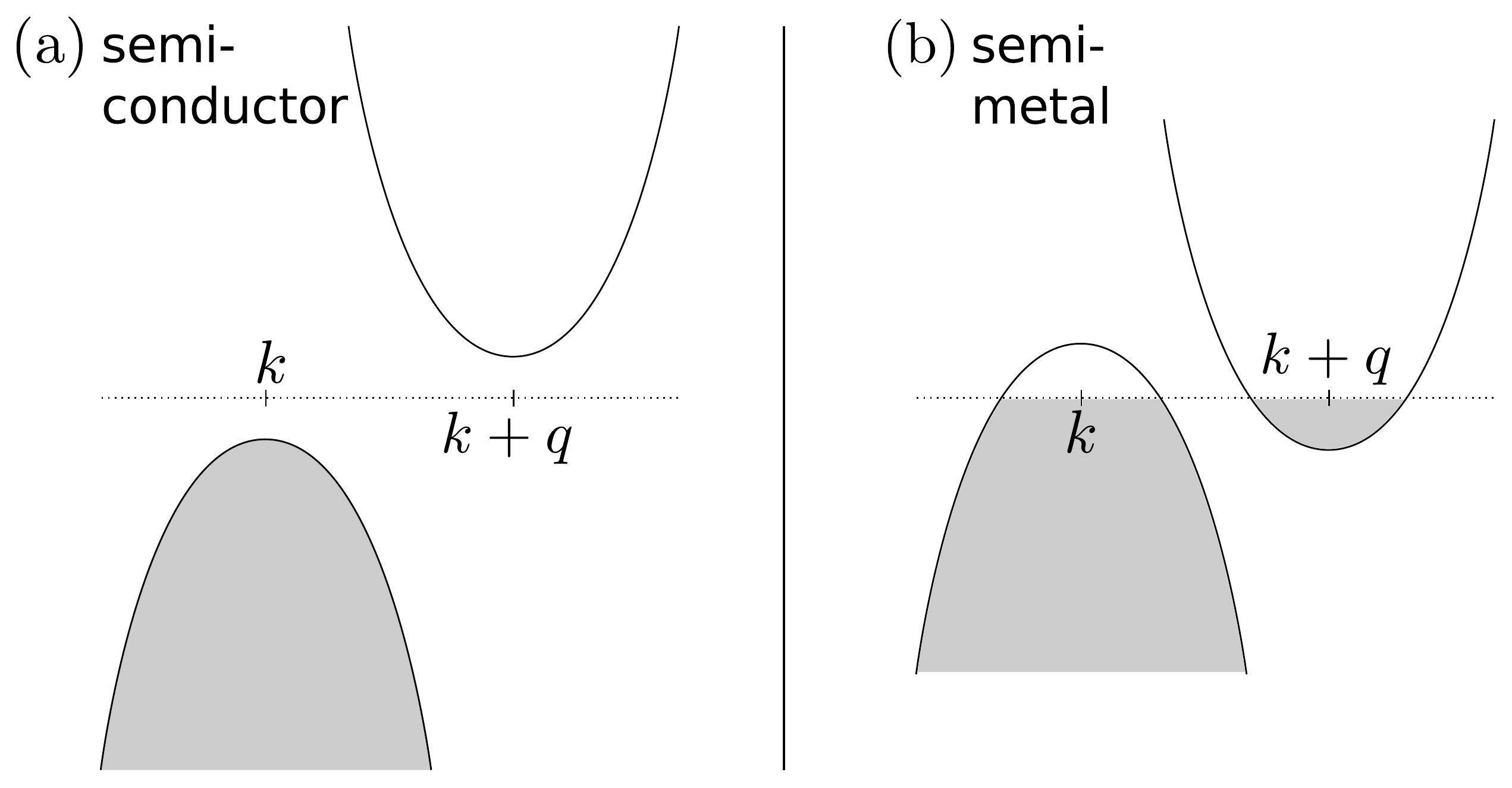}
~~~
\includegraphics[width=0.32\textwidth]{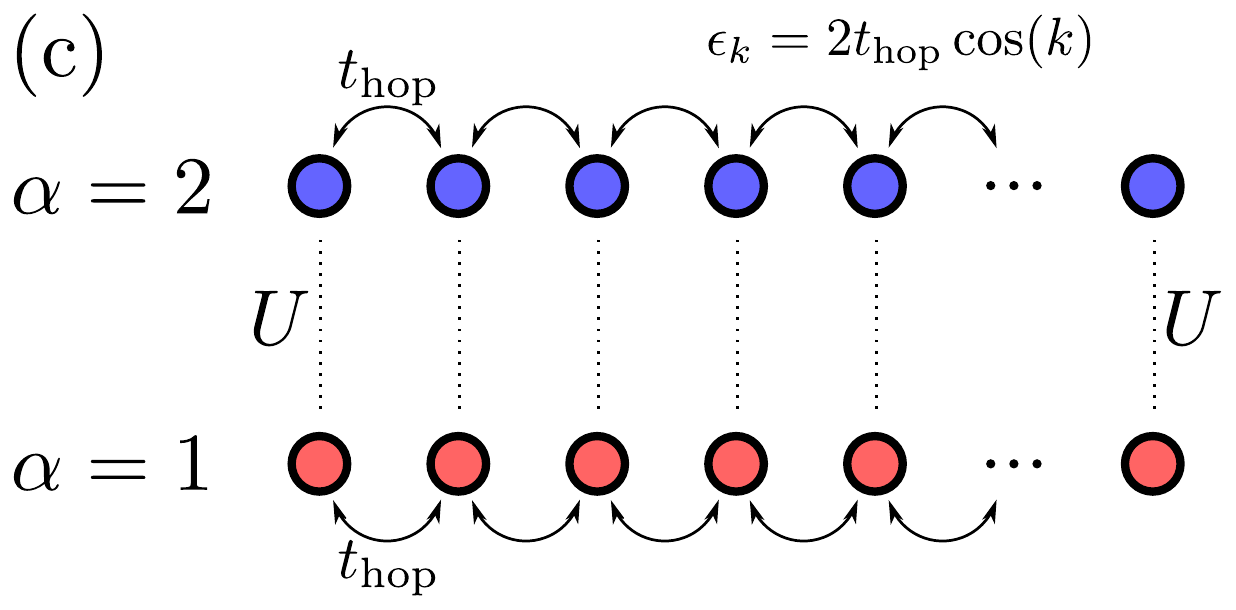}
~~~
\includegraphics[height=0.22\textwidth]{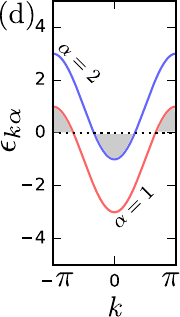}
\caption{(a-b) Schematic energy bands in semiconducting and semimetallic systems. (c) One-dimensional model for the excitonic insulator and (d) the corresponding noninteracting band structure, where $t_{\text{hop}}=-1$ and $\varDelta=2$.}
\label{fig:chains}
\end{figure}

Electron--hole pairs or excitons, bound together by the Coulomb interaction, may spontaneously form in a semiconductor with a narrow energy gap or in a semimetal with a small band overlap, see Fig.~\ref{fig:chains}(a-b). At sufficiently small gaps or overlaps (and low temperatures) compared to the exciton binding energy, the system can become unstable toward an excitonic insulator (EI) phase. The EI, which is based on a purely electronic mechanism, has been proposed already in the sixties~\cite{Mott1961,Keldysh1965,Jerome1967}. In the semi-metal case it is conceptually very similar to BCS superconductivity, where electrons are bound together as Cooper pairs~\cite{BCS1,BCS2,Gorkov1959}. Even if in the original BCS theory the pairing mechanism is due to the electron--phonon interaction, the EI is very interesting to study due to this apparent connection. Moreover, recent works have suggested that the EI phase is realized in materials~\cite{Rossnagel2002,Kogar2017} and can be probed out of thermal equilibrium by time-resolved spectroscopies~\cite{Hellmann2012,Mor2017,Werdehausen2018}, which is our motivation for the present work.

We model the EI by a one-dimensional two-band system with interband Hubbard interaction~\cite{Golez2016}. We can view this as two lattice systems separated in energy and connected by the interaction, see Fig.~\ref{fig:chains}(c). The model Hamiltonian is written in terms of creation, $\hat{c}^\dagger$, and annihilation, $\hat{c}$, operators for spinless electrons:
\be\label{eq:hamiltonian}
\hat{H} = \sum_{ij\a} h_{i_\a j_\a}\hat{c}_{i_\a}^\dagger \hat{c}_{j_\a} + \frac{1}{2}\sum_i U \hat{c}_{i_1}^\dagger \hat{c}_{i_1} \hat{c}_{i_2}^\dagger \hat{c}_{i_2} , 
\ee
where the indices $i,j$ label the lattice sites in the subsystems $\a=\{1,2\}$, see Fig.~\ref{fig:chains}(c). We consider a finite lattice system with $N/2$ sites in each of the subsystems $\a$. The parameters $h_{i_\a j_\a}$ are chosen such that $h_{i_\a j_\a} = t_{\text{hop}}$ for nearest neighbors with periodic boundary conditions. In addition, we choose an on-site energy for the subsystems as $h_{i_\a i_\a} = \varDelta_\a$ with $\varDelta_{1(2)} = \varmp \varDelta/2$. Transforming to $k$-space [see Appendix~\ref{sec:app}] we obtain the well-known energy dispersion for the noninteracting bands $\eps_k = 2t_{\text{hop}}\cos(k)$, where $k$ is discretized as $k=2\pi m/(N/2)$ with $m\in[-N/4, N/4[$. The kinetic part of the Hamiltonian could then be equivalently written as  $\sum_{k\alpha} \eps_{k\alpha}\hat{c}_{k\alpha}^\dagger \hat{c}_{k\alpha}$ where the bands are separated by a direct gap, $\eps_{k\a} = \eps_k+\varDelta_\alpha$, see Fig.~\ref{fig:chains}(d).

The electrons in the upper band are bound to the holes, or repelled by the electrons in the lower band by a local density-density interaction of strength $U$.  More complicated (long-range) interactions are possible to include similarly~\cite{Golez2016}. By the parameter $\varDelta$ we can tune the bands so that there is an electron pocket in the upper band around $k=0$, and a hole pocket in the lower band around $k=\pm\pi$, see Fig.~\ref{fig:chains}(d), and we consider the excitonic pairing of these sectors. Then, for this system to exhibit the EI phase, we consider the density matrix element $\langle \hat{c}_{(k+\pi)1}^\dagger \hat{c}_{k2}\rangle$ to be nonzero; which breaks the conservation of charge within each band and spatial symmetry (charge-density wave). 

We fix $t_{\text{hop}}=-1$ and calculate energies in units of $|t_{\text{hop}}|$ and times in units of $\hbar/|t_{\text{hop}}|$.

\section{Key objects and NEGF equations}\label{sec:background}
In this section we briefly outline the main points in the NEGF theory which are important for the present study. For a more detailed discussion we refer the reader to, e.g., Refs.~\cite{svlbook,Balzerthesis,Balzer2013book}.

To calculate time-dependent nonequilibrium quantities we use the equations of motion for the one-particle Green's function on the Keldysh contour $\g$. This quantity is defined as the ensemble average of the contour-ordered product of particle creation and annihilation operators in the Heisenberg picture~\cite{svlbook}
\be\label{eq:greenf}
G_{i_\a j_\b}(z,z') = -\mathrm{i} \langle {T}_{\gamma} [\hat{c}_{i_\a,\text{H}}(z) \hat{c}_{j_\b,\text{H}}^\dagger(z')] \rangle ,
\ee
where the variables $z$, $z'$ run on the contour. The contour has a forward and a backward branch on the real-time axis, $[t_0,\infty[$, and also a vertical branch on the imaginary axis, $[t_0,t_0-\im\b]$ with inverse temperature $\b$. The Green's function matrix, $G$, with matrix elements defined in Eq.~\eqref{eq:greenf}, satisfies the equation of motion (and the corresponding adjoint equation)~\cite{svlbook}
\be\label{eq:fulleom}
\left[\im\partial_z - h(z)\right]G(z,z') = \delta(z,z') + \int_\g \ud \bar{z} \varSigma(z,\bar{z}) G(\bar{z},z')
\ee
with $\varSigma$ being the self-energy. Depending on the arguments $z,z'$, the Green's function, $G(z,z')$, and the self-energy, $\varSigma(z,z')$, defined on the time contour have components lesser ($<$), greater ($>$), retarded (R), advanced (A), left ($\lceil$), right ($\rceil$) and Matsubara (M)~\cite{svlbook}.

\subsection{Time-stepping procedure}
The Kadanoff--Baym equations (KBE) for the lesser and greater Keldysh components of the Green's function are~\cite{Stan2009}
\begin{align}
\im\partial_t G^\lessgtr(t,t') & = \heff(t) G^\lessgtr(t,t') + I_1^\lessgtr(t,t') \\
-\im\partial_{t'} G^\lessgtr(t,t') & = G^\lessgtr(t,t')\heff(t')  + I_2^\lessgtr(t,t'),
\end{align}
where the effective Hamiltonian is composed of the single-particle Hamiltonian and the time-local Hartree--Fock (HF) self-energy as $\heff(t) \equiv h(t) + \varSigma_{\text{HF}}(t)$. The collision integrals, $I$, incorporate the many-body (beyond HF) self-energies, $\varSigma_{\text{MB}}$. For the following considerations, the exact form of the self-energies is not important, and we will discuss this later in Sec.~\ref{sec:sigma}. By considering only the real-time branch of the Keldysh contour we have, employing the Langreth rules~\cite{svlbook},
\begin{align}
I_1^\lessgtr(t,t') & = \int_{t_0}^t \ud \tb \varSigma_{\text{MB}}^{\text{R}}(t,\tb)G^\lessgtr(\tb,t') \nonumber \\
& + \int_{t_0}^{t'} \ud \tb \varSigma_{\text{MB}}^\lessgtr(t,\tb) G^{\text{A}}(\tb,t') , \\ \nonumber \\
I_2^\lessgtr(t,t') & = \int_{t_0}^t \ud \tb G^{\text{R}}(t,\tb)\varSigma_{\text{MB}}^\lessgtr(\tb,t') \nonumber \\
& + \int_{t_0}^{t'} \ud \tb G^\lessgtr(t,\tb)\varSigma_{\text{MB}}^{\text{A}}(\tb,t') .
\end{align}
From the KBE we obtain for the equal-time limit ($t\to t'^+$)~\cite{Stan2009,Balzer2013book}
\be\label{eq:equalt}
\im\frac{\ud}{\ud t}G^<(t,t) = [\heff(t),G^<(t,t)] + I_{12}^<(t) ,
\ee
where we defined 
\begin{align}
I_{12}^<(t) & \equiv I_1^<(t,t) - I_2^<(t,t) \nonumber\\
& = \int_{t_0}^t \ud \bar{t} \left[\varSigma_{\text{MB}}^>(t,\bar{t})G^<(\tb,t) - \varSigma_{\text{MB}}^<(t,\tb)G^>(\tb,t) \right.\nonumber\\
& \left. \hspace{18pt} + \ G^<(t,\tb)\varSigma_{\text{MB}}^>(\tb,t) - G^>(t,\tb)\varSigma_{\text{MB}}^<(\tb,t)\right], \nonumber \\ \label{eq:collint}
\end{align}
and we used 
\be
k^{\text{R}}(t,t') = \theta(t-t')\left[k^>(t,t') - k^<(t,t')\right] 
\ee
together with the symmetry relation $k^{\text{R}}(t,t') = \left[k^{\text{A}}(t',t)\right]^\dagger$ for $k=G,\varSigma_{\text{MB}}$.

We now explain how to propagate Eq.~\eqref{eq:equalt} from $t\to t+\delta$. The one-particle Hamiltonian is known explicitly as a function of time, so it may be evaluated at half the time-step, and furthermore we introduce~\cite{Stan2009}
\be
U(t) = \ex^{-\im\heffb t} ,
\ee
where $\heffb \equiv h(t+\delta/2) + \varSigma_{\text{HF}}(t)$. In addition, it is useful to introduce a transformation
\be\label{eq:trfo}
G^\lessgtr(t,t') \equiv U(t)\widetilde{G}^\lessgtr(t,t')U^\dagger(t'),
\ee
which incorporates the ``trivial evolution'' due to the effective single particle Hamiltonian. Applying Eq.~\eqref{eq:trfo} in Eq.~\eqref{eq:equalt} and canceling terms leads to
\be\label{eq:appr}
\im\frac{\ud}{\ud t} \widetilde{G}^<(t,t) = U^\dagger(t)I_{12}^<(t)U(t),
\ee
where we approximated $\heff(t) \approx \heffb$. Now, we may integrate over $t$ to obtain
\be
\widetilde{G}^<(t+\delta,t+\delta) = \widetilde{G}^<(t,t) - \im \int_t^{t+\delta} \ud \tb U^\dagger(\tb)I_{12}^<(\tb)U(\tb)
\ee
and using the transformation~\eqref{eq:trfo} again we get
\begin{align}\label{eq:final}
& G^<(t+\delta,t+\delta) \nonumber \\
& = U(t+\delta)\widetilde{G}^<(t,t)U^\dagger(t+\delta) \nonumber \\
& - \im U(t+\delta)\int_t^{t+\delta} \ud \tb U^\dagger(\tb)I_{12}^<(\tb)U(\tb)U^\dagger(t+\delta) \nonumber \\
& = U(\delta)G^<(t,t)U^\dagger(\delta) \nonumber \\
& + U(\delta)\left[-\im\int_0^\delta \ud \tb U^\dagger(\tb)I_{12}^<(\tb+t)U(\tb)\right]U^\dagger(\delta) ,
\end{align}
where we combined the evolution operators using their group property. The integrand has a form for which we may use the Baker--Hausdorff--Campbell expansion
\begin{align}
\ex^A B \ex^{-A} & = B + [A,B] + \frac{1}{2}[A,[A,B]] \nonumber \\
& + \frac{1}{3}[A,\frac{1}{2}[A,[A,B]]] + \ldots,
\end{align}
where $A = \im\heffb \tb$ and $B = I_{12}^<(\tb+t)$. If we assume that the collision integral does not change in the interval $[0,\delta]$, $I_{12}^<(\tb+t) \approx I_{12}^<(t)$, we may perform the integral 
\begin{align}\label{eq:integral}
& -\im\int_0^\delta \ud \tb U^\dagger(\tb)I_{12}^<(\tb+t)U(\tb) \nonumber \\
& \approx -\im \delta I_{12}^<(t) - \frac{\im^2}{2}\delta^2[\heffb,I_{12}^<(t)] \nonumber \\
& - \frac{\im^3}{6}\delta^3[\heffb,[\heffb,I_{12}^<(t)]] \nonumber \\
& - \frac{\im^4}{24}\delta^4[\heffb,[\heffb,[\heffb,I_{12}^<(t)]]] - \ldots .
\end{align}
We may write this in a recursive form by introducing $c_0 = -\im I_{12}^<(t)\delta$ and $c_n = \frac{\im\delta}{n+1}[\heffb,c_{n-1}]$. Finally, the time-diagonal propagation of the lesser Green function is done by inserting Eq.~\eqref{eq:integral} into Eq.~\eqref{eq:final}~\cite{Stan2009,Balzer2013book}.

We summarize the time-stepping procedure on the time-diagonal as the following set of equations
\begin{align}
G^<(t+\delta, t+\delta) & = U(\delta)\left[G^<(t,t) + C\right]U^\dagger(\delta), \label{eq:step} \\
U(\delta) & = \ex^{-\im\heffb \delta}, \\
C & = \sum_{n=0}^\infty c_n(t) , \label{eq:coefsum} \\
c_n(t) & = \frac{\im\delta}{n+1}\left[\heffb,c_{n-1}(t)\right] , \label{eq:coef}\\
c_0(t) & = -\im\delta I_{12}^<(t).
\end{align}
In practice, in Eq.~\eqref{eq:coefsum} we truncate the infinite summation at $n=N_{\text{max}}$ when a desired accuracy is reached for the euclidean norm $||c_{N_{\text{max}}}-c_{N_{\text{max}}-1}||$ from Eq.~\eqref{eq:coef}. Also, as the $n$-th term in the summation is already of the order $\delta^{n+1}$, and as we already approximated $\heff(t)\approx \heffb$ in Eq.~\eqref{eq:appr} and $I_{12}^<(\bar{t}+t)\approx I_{12}^<(t)$ in Eq.~\eqref{eq:integral}, going beyond $N_{\text{max}}=3$ typically does not yield further accuracy.

\subsection{Employing the Generalized Kadanoff--Baym Ansatz}
The GKBA for the greater/lesser Green function is~\cite{Lipavsky1986}
\be\label{eq:gkba}
G^\lessgtr(t,t') \approx \im\left[G^{\text{R}}(t,t')G^\lessgtr(t',t') - G^\lessgtr(t,t)G^{\text{A}}(t,t')\right] .
\ee
Importantly, this still involves double-time propagators $G^{\text{R}/\text{A}}$ which need to be provided for the approximation to be complete. Once this is done, Eq.~\eqref{eq:step} may be used to propagate the lesser Green's function, and the greater component is obtained from the relation $G^>(t,t) = -\im + G^<(t,t)$. We describe the retarded/advanced propagators at the HF level, i.e., we have a bare propagator where the (time-local) HF self-energy is included in the single-particle Hamiltonian $\heff$. Explicitly, the retarded and advanced Green's functions are approximated as~\cite{Balzer2013book}
\begin{align}\label{eq:prop}
G^{\text{R}/\text{A}}(t,t') & \approx \mp \im \theta[\pm(t-t')]{T}\ex^{-\im\int_{t_0}^t\ud \tb \heff(\tb)} \nonumber \\
& \equiv \mp \im \theta[\pm(t-t')]Y(t,t')
\end{align}
where we introduced a ``time-evolution'' operator $Y$ which satisfies $Y(t,t') = [Y(t',t)]^\dagger$ and $Y(t,t) = \unity$. We then insert the GKBA from Eq.~\eqref{eq:gkba} into the collision integral in Eq.~\eqref{eq:collint}. After some simplification and using the introduced operator $Y$ we obtain~\cite{Balzer2013book}
\begin{align}\label{eq:i12}
& I_{12}^<(t) = \nonumber \\
& \int_{t_0}^t  \ud \tb \left\{\left[\varSigma_{\text{MB}}^>(t,\tb)G^<(\tb,\tb) - \varSigma_{\text{MB}}^<(t,\tb)G^>(\tb,\tb) \right]Y(\tb,t)\right. \nonumber \\
& \hspace{10pt} + \ \left. Y(t,\tb)\left[G^<(\tb,\tb)\varSigma_{\text{MB}}^>(\tb,t) - G^>(\tb,\tb)\varSigma_{\text{MB}}^<(\tb,t) \right]\right\} . \nonumber \\
\end{align}
Provided that the time-step length $\delta$ is small we may use a recurrence relation for the time-evolution~\cite{Balzer2013book}
\be
Y(\tb,t) = Y(\tb,t-\delta)U^\dagger(t-\delta)
\ee
which reduces the requirement of diagonalizations of $\heff$ for the evaluation of $Y$. This also means we do not have to worry about the time-ordering ${T}$ in Eq.~\eqref{eq:prop} as the single-particle Hamiltonians (inside the integral) are assumed to be constant in the successive time intervals.

\subsection{Self-energy approximations}\label{sec:sigma}
In practice, we need an approximation for the self-energies discussed in the previous subsections. Also, these quantities are to be represented in some basis, and we choose the localized \emph{site} basis of our EI system, see Fig.~\ref{fig:chains}(c); this means that the system is described as a lattice with basis functions describing localized orbitals around the lattice sites. In the following we refer to the lattice sites with latin indices ($i,j,k,l$), and to the separate subsystems with greek indices ($\a,\b,\g,\d$).

For the many-body self-energy we take the ``second-Born approximation'' (2B)~\cite{Hermanns2012,Latini2014}. The HF and 2B self-energies are:
\begin{align}
(\varSigma_{\text{HF}})_{i_\a j_\b}(t) & = \xi\delta_{ij}\delta_{\a\b}\sum_{k\g} v_{i_\a k_\g}(t)[-\im G_{k_\g k_\g}(t,t)] \nonumber \\
& - v_{i_\a j_\b}(t) [-\im G_{j_\b i_\a}(t,t)], \label{eq:hfs}\\ \nonumber \\
(\varSigma_{\text{MB}})_{i_\a j_\b}(t,t') & = \sum_{\substack{kl \\ \g\d}} v_{i_\a k_\g}(t) v_{j_\b l_\d}(t') G_{k_\g l_\d}(t',t) \nonumber \\
& \times \left[\xi G_{j_\b i_\a}(t,t')  G_{l_\d k_\g}(t,t') \right.\nonumber \\
& \left.\hspace{1pt} - \ G_{j_\b k_\g}(t,t') G_{l_\d i_\a}(t,t') \right] , \label{eq:2bs}
\end{align}
with a spin-degeneracy factor $\xi$ for the direct terms~\cite{Balzer2013book}. Our model is for spinless fermions, such that $\xi=1$. For the 2B self-energy (due to being non-local in time) it is then desirable to use the GKBA for the Green's function entries. Also, we only need the lesser/greater components to be inserted in Eq.~\eqref{eq:i12}, and by employing Eqs.~\eqref{eq:gkba} and~\eqref{eq:prop} these become
\begin{align}\label{eq:s2b}
& (\varSigma_{\text{MB}})_{i_\a j_\b}^\lessgtr(t,t') \nonumber \\
& = \sum_{\substack{kl \\ \g\d}}v_{i_\a k_\g}(t)v_{j_\b l_\d}(t')[G^\gtrless(t',t')Y(t',t)]_{l_\d k_\g} \nonumber \\
& \times \left\{ \xi [Y(t,t')G^\lessgtr(t',t')]_{i_\a j_\b}[Y(t,t')G^\lessgtr(t',t')]_{k_\g l_\d} \right.\nonumber \\ 
& \left.\hspace{1pt}- \ [Y(t,t')G^\lessgtr(t',t')]_{i_\a l_\d}[Y(t,t')G^\lessgtr(t',t')]_{k_\g j_\b}\right\}. \nonumber \\
\end{align}

Even though the 2B approximation goes beyond the effective one-particle description of HF, it still includes only bare interaction up to second order, i.e., it neglects screening effects and higher order correlations, such as in the $GW$~\cite{Hedin1965} or $T$-matrix~\cite{Galitskii1958} approximations. However, compared to the full two-time KBE, only the 2B approximation together with the GKBA allows for a maximal speed-up in computational scaling ($\sim T^2$ versus $\sim T^3$, $T$ being the total propagation time) due to the higher scaling of these more accurate approximations.

\section{Correlated equilibrium state}\label{sec:results}

\subsection{Initial preparation by a HF iteration}

Since the HF self-energy is local in time, we may perform a simple time-independent calculation to obtain the HF density matrix. In this procedure, we simply solve the eigenvalue problem for the effective Hamiltonian~\cite{Roothaan1951,Hall1951,Balzer2009}
\be
( h + \varSigma_{\text{HF}} ) \ket{\psi} = \vareps \ket{\psi} . 
\ee
This is an iterative process where (1) an initial density matrix is given; (2) the HF self-energy is constructed from the given density matrix; (3) the effective hamiltonian is constructed from the HF self-energy and the corresponding eigenvalue problem is solved; (4) a new density matrix is constructed from the eigenvectors of step 3:
\be
\rho = \sum_j f(\vareps_j)\ket{\psi_j} \bra{\psi_j},
\ee
where $f(\vareps_j) = [\ex^{\beta(\vareps_j - \mu)}+1]^{-1}$ is the Fermi function at inverse temperature $\b$ and chemical potential $\mu$. In practice we consider half-filling and choose the chemical potential between the two centermost eigenvalues. This density matrix is then used again for calculating a new HF self-energy in step 2. It is also customary to iteratively mix the old and new density matrices as $\a \rho_{\text{new}} + (1-\a)\rho_{\text{old}}$ with $\a$ a real number between $0$ and $1$.

In our EI model, we are considering a symmetry-broken ground state where the density matrix has off-diagonal elements related to the exciton pairing. If we start the above-mentioned iteration procedure from a purely noninteracting initial density matrix, there is no way for the iteration to gain nonzero off-diagonal elements. To go around this, we introduce a weak coupling between the subsystems, and use this as a `seed state' which has a physical nonzero contribution to the off-diagonal parts of the density matrix. Once the HF iteration has converged we have an excitonic state. This state is then used as a `seed state' for another HF iteration with a weaker coupling between the subsystems. This procedure is continued by weakening the coupling at every stage of the iteration, until we reach zero coupling between the subsystems. This is then the true physical setting in our model, and we have a convergence to an excitonic state, provided that the system parameters ($U,\varDelta,\beta$) are favoring this.

\subsection{Solving the Dyson equation on the imaginary-time contour}
For general time coordinates $z,z'$ on the full complex-time contour the equation of motion for the Green's function is in Eq.~\eqref{eq:fulleom}. On the vertical branch of the time contour we have (we assume $t_0=0$) $z=-\im\tau$ and $\tau \in [0,\beta]$; $\delta(z,z') = \im\delta(\tau-\tau')$. Also, the system is time-independent, so $h(z) = h$, and the Green's function and self-energy depend on the time difference only: $G^{\text{M}}(\tau-\tau') \equiv -\im G(-\im\tau,-\im\tau')$ and $\varSigma^{\text{M}}(\tau-\tau') \equiv -\im \varSigma(-\im\tau,-\im\tau')$. The equation of motion then takes the form~\cite{Dahlen2005,Balzer2009}
\begin{align}\label{eq:eomtau}
& [-\partial_\tau - h]G^{\text{M}}(\tau-\tau') \nonumber \\
& = \delta(\tau-\tau') + \int_0^\beta \ud \bar{\tau} \varSigma^{\text{M}}(\tau-\bar{\tau}) G^{\text{M}}(\bar{\tau}-\tau'). 
\end{align}
For practical purposes it is convenient to consider a change of variable to $\tau-\tau'\equiv \widetilde{\tau}\in[-\beta,\beta]$, and use the fact that both the Green's function and self-energy are $\beta$-anti-periodic. In the end, only one half of the range in $\widetilde{\tau}$ is needed and it is convenient to choose $\widetilde{\tau}\in[-\b,0]$ since the initial density matrix is constructed from $G^{\text{M}}(0^-)$.

Eq.~\eqref{eq:eomtau} is transformed into an integral equation by introducing a reference Green's function $G_0^{\text{M}}$ satisfying $[-\partial_{\widetilde{\tau}} - h - \varSigma_0^{\text{M}}]G_0^{\text{M}}(\widetilde{\tau}) = \delta(\widetilde{\tau})$, where $\varSigma_0^{\text{M}}$ is the local part of the self-energy: $\varSigma^{\text{M}}(\tau) = \varSigma_0^{\text{M}}\delta(\tau) + \varSigma_c^{\text{M}}(\tau)$. The integral form reads~\cite{Balzer2009,Uimonenthesis}
\begin{align}\label{eq:integraltau}
& G^{\text{M}}(\widetilde{\tau}) - G_0^{\text{M}}(\widetilde{\tau}) = \nonumber \\
& - \int_0^\b \!\! \ud \tau_1 \int_0^\b \!\! \ud \tau_2 G_0^{\text{M}}(\widetilde{\tau}-(\tau_1-\b))\varSigma_c^{\text{M}}(\tau_1-\tau_2) G^{\text{M}}(\tau_2) , \nonumber \\
\end{align}
where the nonlocal correlations are included in $\varSigma_c^{\text{M}}$. Eq.~\eqref{eq:integraltau} is typically further rewritten as a Fredhold integeral equation~\cite{Balzer2009,Uimonenthesis,arfkenweber}
\be\label{eq:axb}
\int_{-\beta}^0 \ud \tau' A(\tau,\tau') G^{\text{M}}(\tau') = G_0^{\text{M}}(\tau)
\ee
by introducing $A(\tau,\tau') \equiv \delta(\tau-\tau') - F(\tau,\tau')$ and $F(\tau,\tau') \equiv \int_0^\beta \ud \tau_1 G_0^{\text{M}}(\tau-(\tau_1-\beta))\varSigma_c^{\text{M}}(\tau_1-(\tau'+\b))$. Effectively, in Eq.~\eqref{eq:axb}, we are then left with an ``$Ax=b$'' set of linear equations where $A$ consists of the Fredholm integral kernel, $x$ is the (unknown) Matsubara Green's function, and $b$ is the reference Green's function. Typically, the reference Green's function, $G_0^{\text{M}}$, is convenient to construct from the HF solution from the previous subsection. The observables obtained from the self-consistent $G^{\text{M}}$, however, should not dependent on the choice of the reference $G_0^{\text{M}}$, see Eq.~\eqref{eq:eomtau}~\cite{Dahlen2005}. At the second-Born correlations level we would then use $\varSigma_0^{\text{M}} = \varSigma_{\text{HF}}$ and $\varSigma_{c}^{\text{M}}=\varSigma_{\text{MB}}$ from Eqs.~\eqref{eq:hfs} and~\eqref{eq:2bs} for time-arguments on the vertical branch of the time-contour~\cite{Dahlen2005,Balzer2009,Schueler2018}.

In the following, we refer to the solution of the Dyson equation on the imaginary-time contour simply as the ``Matsubara calculation''.

\begin{figure*}[htb]%
\includegraphics*[width=\textwidth]{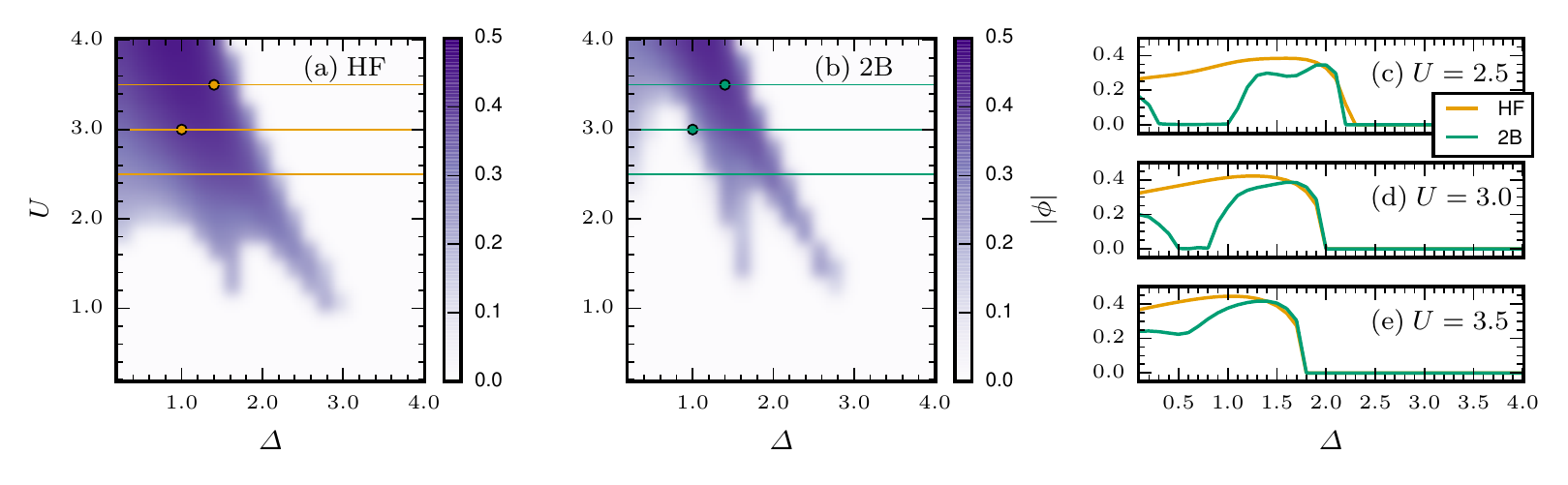}
\caption{Equilibrium phase diagrams of the EI system evaluated by (a) time-independent HF iteration and (b) solving the imaginary-time Dyson equation using the 2B self-energy. Panels (c)-(e) show the excitonic order parameter versus the energy gap for a fixed $U=\{2.5,3.0,3.5\}$ indicated by the horizontal lines in panels (a) and (b). The markers in panels (a) and (b) correspond to the simulations in Figs.~\ref{fig:switch} and~\ref{fig:switchfail}.}
\label{fig:phase}
\end{figure*}

\subsection{Adiabatic preparation of a correlated state}
We can now investigate how the correlated equilibrium state can be prepared at the 2B level. We first perform two separate calculations according to the previous subsections, a time-independent HF calculation and a Matsubara calculation using the 2B self-energy. From this comparison we see how far are the HF and 2B solutions from each other; this is important because we wish to adiabatically evolve from the HF solution into the 2B solution.  Even though the HF and Matsubara calculations can be performed at a finite temperature $1/\b$, we wish to consider effectively a zero-temperature limit ($\b=100$) as the AS procedure is consistent only at zero temperature.

We characterize the EI phase by momentum-averaging the excitonic order parameter over the reduced Brillouin zone (RBZ)~\cite{Golez2016}
\be\label{eq:eiop}
\phi = |\phi|\ex^{\im\theta} \equiv \frac{1}{N_k}\sum_{{k\in[-\frac{\pi}{2},\frac{\pi}{2}[}}\langle \hat{c}_{(k+\pi)1}^\dagger \hat{c}_{k2}\rangle,
\ee
where $N_k$ is the number of $k$ points in the RBZ. We discuss the details in Appendix~\ref{sec:app} on how to extract this quantity from our localized site basis representation of the density matrix. In addition to the excitonic order parameter we consider the total energy~\cite{Dahlen2005,Balzer2013book}
\be
E_{\text{tot}} = E_0 + E_{\text{HF}} + E_{\text{correlation}},
\ee
where $E_0 = \Re\text{Tr}[h_0 \rho]$, $E_{\text{HF}} = \frac{1}{2}\Re\text{Tr}[\varSigma_{\text{HF}} \rho]$, and $E_{\text{correlation}} = -\frac{1}{2}\Im \text{Tr}[I_{12}^>]$, $h_0$ being the kinetic part of the Hamiltonian in Eq.~\eqref{eq:hamiltonian} and $\rho=-\im G^<$.

In Fig.~\ref{fig:phase}, for a system of $N=24$ sites at $\b=100$, we show the equilibrium phase diagrams where the absolute value of the (complex) excitonic order parameter is plotted against the energy gap $\varDelta$ and the interaction strength $U$. We notice a general trend that for smaller gaps the system behaves as a normal semimetal whereas when the gap is larger the system goes towards normal semiconducting and insulating states. Between these two regimes the system exhibits the symmetry-broken EI phase, when the interaction strength is suitable for pair formation. By looking at fixed-$U$-lines in Fig.~\ref{fig:phase}(c)-(e) we can see a typical behavior of the excitonic order parameter versus the energy gap: We see a sharp drop to a semiconducting or insulating state at a critical value for $\varDelta$ which could then be related to the exciton binding energy. For small $\varDelta$ the decay to a semimetallic state is slower, cf.~\cite{Jerome1967}. In Fig.~\ref{fig:phase}(b) we see that the 2B approximation retains the overall feature of the HF phase diagram, but the range in $\varDelta$ and $U$ for which the excitonic order is stabilized is more narrow. In addition, more advanced approximations for correlations, in general, reduce the absolute value for $\phi$ see Fig.~\ref{fig:phase}(c)-(e) and Ref.~\cite{Golez2016}.

For the numerics we point out that the choice of $N=24$ lattice sites is simply for the ease of computation, and here it is justified as we are comparing calculations within the same basis representations, even if the $k$-resolved quantities would not be completely converged in the number of lattice sites. For comparison with a $k$-space calculation~\cite{Golez2016} we have checked that $N=64$ is roughly in agreement (relative error in $n\lesssim 10^{-4}$), but for the purpose of the present analysis this larger basis is not necessary. (The sharp features in Fig.~\ref{fig:phase} possibly result from finite-size effects, and a smoother behavior might be observed with larger $N$.) For the imaginary time grid $[-\b,0]$ we use a uniform power discretization due to the exponential behavior of the Matsubara Green's functions and self-energies around the endpoints~\cite{Ku2002,Dahlen2005,Schueler2018}. The number of grid points in this uniform power mesh is $2up+1$ and we use $u=5$, $p=7$ to achieve a reasonable convergence in total energies: $|1 - E_{\text{tot}}^{(u,p)=(5,7)}/E_{\text{tot}}^{(u,p)=(6,8)} | \lesssim 10^{-6}$.

In the AS procedure, we employ a ramp function in the 2B self-energy in Eq.~\eqref{eq:2bs} for the interaction strength $v(t) = f(t) v_0$ where $v_0$ is the part of Eq.~\eqref{eq:hamiltonian} corresponding to the two-body interaction~\cite{Balzer2013book}. The two-body interaction in the HF self-energy in Eq.~\eqref{eq:hfs} remains static during this procedure. For the ramp function $f$ we choose a double-exponential form, see Refs.~\cite{Watanabe1990,Schluenzen2016}.

\begin{figure}[t]
\centering
\includegraphics[width=0.45\textwidth]{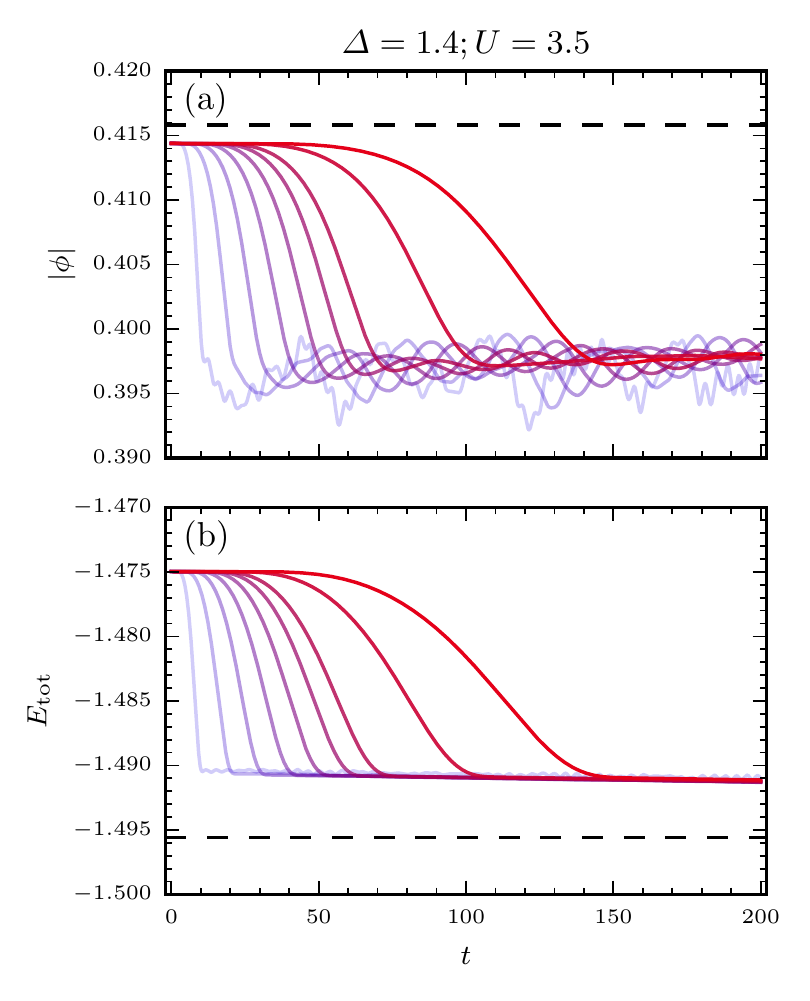}
\caption{Adiabatic switching of the 2B self-energy with $\varDelta=1.4$ and $U=3.5$. The color range from blue (light) to red (dark) indicates increasing switching times from $0.03T$ to $0.42T$ in terms of the total simulation time $T=200$. (a) Excitonic order parameter; (b) total energy. The dashed lines correspond to the equilibrium values from the Matsubara calculation.}
\label{fig:switch}
\end{figure}

\begin{figure}[t]
\centering
\includegraphics[width=0.45\textwidth]{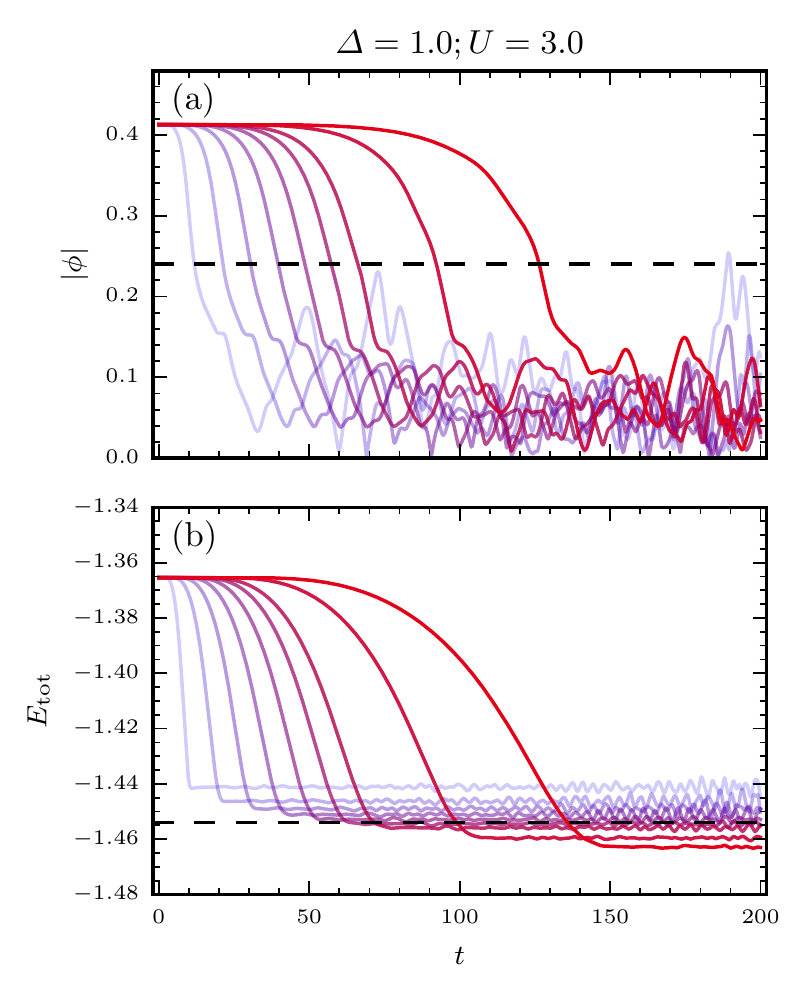}
\caption{Same as Fig.~\ref{fig:switch} but with $\varDelta=1.0$ and $U=3.0$.}
\label{fig:switchfail}
\end{figure}

In Fig.~\ref{fig:switch}(a) we show a propagation with $\varDelta=1.4$ and $U=3.5$ where the excitonic order parameter is reasonably similar and nonzero for both HF and 2B [see Fig.~\ref{fig:phase}(e)]. We see that if the switching is performed too fast, the order parameter has a persistent oscillation, whereas for slower switching the procedure indeed follows an adiabatic behavior, and the order parameter saturates to a roughly fixed value. This value is not exactly the same as from the Matsubara calculation since we lose some information about quasiparticle renormalization due to correlation effects by the HF propagators within the GKBA, see Eq.~\eqref{eq:prop}. Fig.~\ref{fig:switch}(b) shows the same calculation for the total energy, which is a bit more robust regarding its saturation. The result is still reasonable as we have prepared a correlated symmetry-broken initial state by the AS procedure, although it can take relatively long times to saturate. 

In Fig.~\ref{fig:switchfail}(a) we show a propagation using $\varDelta=1.0$ and $U=3.0$. Looking at Fig.~\ref{fig:phase} we see that the HF solution suggests a stronger EI state, whereas in the more correlated 2B approximation it is not as pronounced. This leads to a failure in the AS procedure where the order parameter oscillates persistently and decays towards zero, no matter how slow the switching procedure is (at least within this time window). We note that the total energy in Fig.~\ref{fig:switchfail}(b) shows a more saturated result although it might then be more reasonable to relate this to the energy of the normal state instead of the symmetry-broken EI phase. However, we observe that the oscillations for the slowest switching are away from $|\phi|=0$, so it is plausible that even slower switching procedure might lead to a saturated result corresponding to an EI state.

We point out that the absolute value of the excitonic order parameter decreases during the AS procedure. This happens also in the cases where the 2B value from the Matsubara calculation is higher than the HF value, see Fig.~\ref{fig:switch}. This could be a consequence of the AS procedure itself, or that the 2B value from the GKBA is simply lower than the 2B value from the Matsubara calculation. One could further analyze this by performing a full KB simulation without the imaginary-time branch but with an adiabatic switch-on of the interactions~\cite{Rios2011,Schluenzen2016,Balzer2016}; this is however beyond the scope of the present work. 

\begin{figure}[t]
\centering
\includegraphics[width=0.45\textwidth]{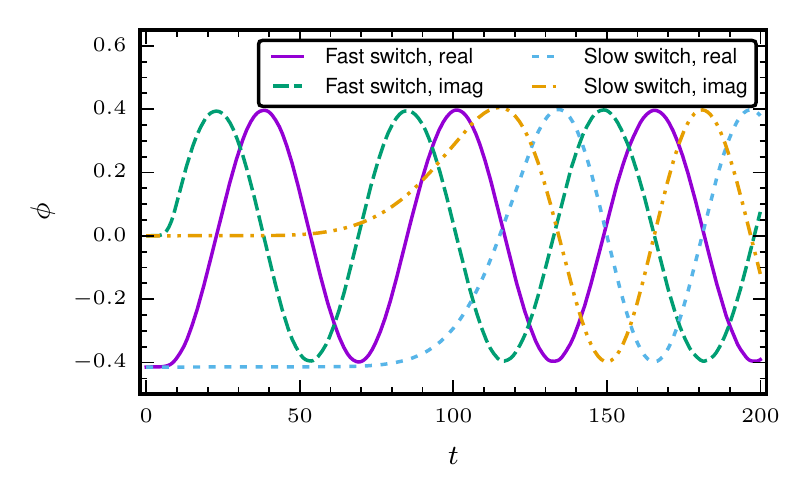}
\caption{Real and imaginary parts of the order parameter in Fig.~\ref{fig:switch} with the fastest ($0.03T$) and slowest ($0.42T$) switching times, respectively.}
\label{fig:oposc}
\end{figure}

Closer inspection also shows that the adiabatic switching procedure generates a nonzero phase $\theta$ of the complex order parameter $\phi$ [Eq.~\eqref{eq:eiop}]. In Fig.~\ref{fig:oposc} we show temporal oscillations of the order parameter's real and imaginary parts, which in the case where the adiabatic switching procedure works are almost perfectly phase-shifted to yield a practically time-independent absolute value $|\phi|$. We note that the observed oscillations are reminiscent of the phase (Nambu--Goldstone)~\cite{Nambu1960,Goldstone1961} and amplitude (Anderson--Higgs)~\cite{Anderson1958,Higgs1964} modes arising in systems with complex order parameters, but caution that their excitation mechanism is the non-physical adiabatic switching of $v(t)$ in $\varSigma_{\text{MB}}$ [Eq.~\eqref{eq:s2b}] in our case.

\section{Conclusions and outlook}\label{sec:concl}

We considered the time-propagation of the nonequilibrium Green's function within the GKBA, to study the symmetry-broken ground state of an excitonic insulator. By comparison to the solution of the Dyson equation on the imaginary branch of the Keldysh contour, the commonly used adiabatic preparation for a correlated initial state by the GKBA was benchmarked. We found that it is possible to prepare a symmetry-broken initial state by the AS procedure although it may take considerably long times to saturate. We expect this behavior to be general for other symmetry-broken or ordered states as well, such as superconducting~\cite{Kemper2015,Sentef2016,Babadi2017,Murakami2017sc} or charge-density wave order~\cite{Shen2014,Huber2014,Schueler2018cdw}. We note that the AS procedure may be problematic if the starting point, in our case the HF initial state, does not describe the state of the system sufficiently well.
Very recently the inclusion of initial correlations within the GKBA has been proposed~\cite{Karlsson2018} which might prove helpful also for symmetry-broken initial states.

The description of the propagators at the HF level might also prevent the system from relaxing due to lack of damping. Nonhermitian contributions for more correlated approximations for the propagators have been discussed, e.g., in Refs.~\cite{Bonitz1996,Jahnke1997,Bonitz1999,Pal2009,Latini2014} but we expect the overall behavior of long saturation be present for more correlated approximations for the propagators as well. In addition, the conservation laws within GKBA at the HF level~\cite{Hermanns2014} might be violated if the quasiparticle contributions are not dealt with self-consistently.

Here we considered a periodic lattice system for which a solution of the KB and GKBA equations would be also possible directly in $k$-space. Our implementation in the localized lattice site basis has been tested to be in agreement with a $k$-space calculation, but for future studies the site basis implementation readily allows us to consider also disordered systems breaking the lattice periodicity, or real-time charge and thermal transport setups with lead environments~\cite{Jauho1994,Myohanen2009,Latini2014,Tuovinen2014,Karlsson2014,Covito2018}.

For a properly prepared correlated symmetry-broken initial state the next steps include out-of-equilibrium simulations in a pump--probe setting~\cite{Eckstein2008,Freericks2009,Sentef2013,Sentef2013b,Kemper2014,Sentef2015,Rameau2016,Wang2017,Topp2018}. Recent time-domain ARPES experiments~\cite{Mor2017} and simulations~\cite{Golez2016,Murakami2017} show both light-induced enhancement and melting of excitonic order. Using the time-propagation based on the GKBA allows further investigation for longer times, especially mapping out nonthermal critical behavior. Similarly, the extension of GKBA to electron--boson systems is of high interest in order to address questions of light-enhanced electron--phonon couplings~\cite{Pomarico2017,Kennes2017,Sentef2017}, quantum nonlinear phononics~\cite{Puviani2018}, or coupling to quantum photons in cavity quantum-electrodynamical materials science~\cite{Sentef2018,Schlawin2018,Mazza2018,Curtis2018}.

\begin{acknowledgement}
R.T. and M.A.S. acknowledge funding by the DFG (Grant No. SE 2558/2-1) through the Emmy Noether program. D.G. and M.S. were supported by the Swiss National Science Foundation through Grant No. 200021\_165539 and ERC Consolidator Grant 724103. M.E. acknowledges funding by the ERC Starting Grant 716648. We wish to thank Daniel Karlsson for productive discussions.
\end{acknowledgement}

\appendix
\section{Basis transformations}\label{sec:app}
In order to evaluate Eq.~\eqref{eq:eiop} from our localized site basis representation of the density matrix, we transform the field operators as
\begin{align}
\hat{c}_{k\a}^\dagger &= \frac{1}{\sqrt{N_\a}}\sum_{i_\a} \ex^{\im k i_\a} \hat{c}_{i_\a}^\dagger, \label{eq:trfo1} \\
\hat{c}_{k\a} &= \frac{1}{\sqrt{N_\a}}\sum_{i_\a} \ex^{-\im k i_\a} \hat{c}_{i_\a}. \label{eq:trfo2}
\end{align}
We label the lattice sites as $i_\a\in\{1,2,\ldots,N/2\}$ for $\a=1$ and $i_\a\in\{N/2+1,N/2+2,\ldots,N\}$ for $\a=2$, thereby giving $N_\a=N/2$ where $N$ is the total number of sites. The summand in Eq.~\eqref{eq:eiop} is then transformed as
\begin{align}
& \langle \hat{c}_{(k+\pi)1}^\dagger \hat{c}_{k2}\rangle \nonumber \\
& = \frac{1}{\sqrt{N/2}} \sum_{i=1}^{N/2} \ex^{\im (k+\pi) i} \frac{1}{\sqrt{N/2}} \sum_{j=N/2+1}^{N} \ex^{-\im k (j-N/2)} \langle \hat{c}_i^\dagger \hat{c}_j \rangle \nonumber \\
& = \frac{2}{N} \sum_{i,j=1}^{N/2} (-1)^i\ex^{\im k (i-j)} \langle \hat{c}_i^\dagger \hat{c}_{j+N/2} \rangle,
\end{align}
where the alternating sign comes from $(\ex^{\im\pi})^i$. Then averaging over the $k$ points in the RBZ gives for Eq.~\eqref{eq:eiop}
\be\label{eq:eiop2}
\phi = \frac{2}{N} \sum_{i,j=1}^{N/2} (-1)^i f_{ij} \langle \hat{c}_i^\dagger \hat{c}_{j+N/2} \rangle ,
\ee
where we introduced
\be\label{eq:ksum}
f_{ij} = \frac{1}{N_k}\sum_{k\in[-\frac{\pi}{2},\frac{\pi}{2}[} \ex^{\im k (i-j)}.
\ee
As we have $N/2$ lattice points corresponding to either one of the full-range bands in Fig.~\ref{fig:chains}(d), for consistency we then choose the number of $k$ points in Eq.~\eqref{eq:ksum} in the RBZ to be $N_k = N/4$. In practice we evaluate the sum numerically but in the limit of infinite number of sites, the behavior of the transformation can be seen from
\be
f_{ij} \stackrel{N\to \infty}{\longrightarrow} \int_{-\pi/2}^{\pi/2} \frac{\ud k}{\pi} \ex^{\im k (i-j)} = \frac{\sin [\frac{\pi}{2} (i-j)]}{\frac{\pi}{2} (i-j)}.
\ee
The excitonic order parameter can now be directly evaluated from Eq.~\eqref{eq:eiop2} since we have our lattice site basis density matrix from the lesser Green's function $\rho=\langle\hat{c}^\dagger \hat{c}\rangle = -\im G^<$. Even though we consider here only the order parameter, we note that other characteristics of the system, such as the band populations, could be evaluated in a similar manner.

%

\providecommand{\WileyBibTextsc}{}
\let\textsc\WileyBibTextsc
\providecommand{\othercit}{}
\providecommand{\jr}[1]{#1}
\providecommand{\etal}{~et~al.}

%



\end{document}